\pdfoutput=1

\documentclass[12pt]{article}
\usepackage[sort,longnamesfirst]{natbib}

\usepackage{amsbsy,amsmath,amsthm,amssymb,graphicx}
\usepackage{multirow}
\usepackage[usenames,dvipsnames]{color}
\usepackage{subfigure}
\usepackage{geometry}
\graphicspath{{./images_jpg/}}

\begin{document}

\title{Emulating a gravity model to infer the spatiotemporal
dynamics of an infectious disease}

\author{Roman Jandarov\\Department of Statistics
  \\ The Pennsylvania State University \\ {\tt raj153@psu.edu}
\and Murali Haran\\
Department of Statistics\\ The Pennsylvania State University \\
{\tt mharan@stat.psu.edu}
\and Ottar Bj{\o}rnstad\\
Departments of  Entomology and Biology\\ The Pennsylvania State
University \\ {\tt onb1@psu.edu}
\and Bryan Grenfell\\
Departments of Ecology and Evolutionary Biology\\ Princeton
University \\ {\tt grenfell@princeton.edu} }

\date{Draft: \today}

\maketitle

\begin{abstract}

  Probabilistic models for infectious disease dynamics are useful for
  understanding the mechanism underlying the spread of infection. When
  the likelihood function for these models is expensive to evaluate,
  traditional likelihood-based inference may be computationally
  intractable. Furthermore, traditional inference may lead to poor
  parameter estimates and the fitted model may not capture important
  biological characteristics of the observed data. We propose a novel
  approach for resolving these issues that is inspired by recent work
  in emulation and calibration for complex computer models. Our
  motivating example is the gravity time series
  susceptible-infected-recovered (TSIR) model. Our approach focuses on
  the characteristics of the process that are of scientific
  interest. We find a Gaussian process approximation to the gravity
  model using key summary statistics obtained from model
  simulations. We demonstrate via simulated examples that the new
  approach is computationally expedient, provides accurate parameter
  inference, and results in a good model fit. We apply our method to analyze measles outbreaks in England
and Wales in two periods, the pre-vaccination period from
1944-1965 and the vaccination period from 1966-1994. Based on our
results, we are able to obtain important scientific insights about
the transmission of measles. In general, our method is applicable
to problems where traditional likelihood-based inference is
computationally intractable or produces a poor model fit. It is
also an alternative to approximate Bayesian computation (ABC) when
simulations from the model are expensive.

\end{abstract}

\section{Introduction} \label{sec:intro}

Infectious disease dynamics are of interest to modelers from a
range of disciplines. The theory of disease dynamics provides a
tractable system for investigating key questions in population and
evolutionary biology. Understanding the disease dynamics helps in
management and with pressing disease issues such as disease
emergence and epidemic control strategies. Probabilistic models
for disease dynamics are important as they help increase our
understanding of the mechanism underlying the spread of the
infection while also accounting for their inherent stochasticity.
Observations on reported cases of the diseases, especially in the
form of space-time data, are becoming increasingly available,
allowing for statistical inference for unknown parameters of these
models. However, traditional likelihood-based inference for many
disease dynamics models is often challenging because the
likelihood function may be expensive to evaluate, making
likelihood-based inference computationally intractable.
Furthermore, traditional inference may lead to poor parameter
estimates and the fitted model may not capture important
biological characteristics of the observed data. Hence, an
approach that simultaneously addresses the computational
challenges as well as the inferential issues would be very useful
for a number of interesting and important probabilistic models for
dynamics of diseases. Inspired by work in the field of emulation
and calibration for complex computer models
\cite[cf.][]{sacks1989design, bayarri2007computer,
craig2001bayesian, kennedy2001bayesian}, we develop a novel
approach for inference for such models. Our approach uses a
Gaussian process approximation to the disease dynamics model using
key biologically relevant summary statistics obtained from
simulations of the model at differing parameter values. As we will
demonstrate, this approach results in reliable parameter estimates
and a good model fit, and is also computationally efficient.

The motivating example for our approach is the gravity time series
susceptible-infected-recovered (TSIR) model for measles dynamics.
The spatiotemporal dynamics of measles have received a lot of
attention in part due to the importance of the disease, the highly
nonlinear outbreak dynamics and also because of the availability
of rich data sets. Important aspects of local dynamics of measles
are well studied. These include key issues like seasonality in
transmission of the infection \citep{dietz1976incidence,
bjornstad2002dynamics}, effects of host demography on outbreak
frequency \citep{mclean1988measles, finkenstadt1998patterns}, and
causes of local persistence and extinctions \citep{Bart:dete:1956,
grenfell1997meta, grenfell2001travelling}. During the course of
outbreaks in well-mixed local populations, the epidemic trajectory
of measles is virtually unaffected by infection that may enter
from neighboring locations. However, spatial coupling is
fundamental to the dynamics and management of measles for smaller
communities where the infection may become locally extinct
\citep{grenfell1997meta, Bart:dete:1956}. Hence, ecologists have
also studied the spatial spread of the disease using so-called
metapopulation models \citep{grenfell1997meta,
swinton1998persistence, earn1998persistence}.

In this paper, we investigate inference for a model first proposed
by \cite{xia2004measles}. The model represents a combination of
the TSIR model \citep{bjornstad2002dynamics, grenfell2002dynamics}
with a term that allows for spatial transmission between different
host communities modeled as a gravity process.
\cite{xia2004measles} demonstrate how this model captures
scientifically important properties of measles dynamics. Since
each likelihood evaluation is computationally very expensive,
however, \cite{xia2004measles} obtain only point estimates of the
parameters minimizing \emph{ad hoc} objective functions instead of
using a likelihood-based approach. Here, we develop a more
statistically rigorous approach to inferring model parameters,
characterizing associated uncertainties and carefully studying
parameter identifiability issues. First, in order to explain the
issues that arise in inferring these parameters via a
likelihood-based approach, we propose a partial discretization of
the parameter space that allows us to perform Bayesian inference
for the parameters using a fast MCMC algorithm. Using this
approach we are able to study uncertainties about the parameter
values. The method allows us to investigate parameter
identifiability issues, showing which gravity model parameters can
or cannot be inferred from a given data set. However, this
approach to resolving the computational challenges of traditional
likelihood-based inference is problematic, as is revealed by our
simulated data examples. We find that the parameter estimates are
poor and the forward simulations of the model at these parameter
settings do not reproduce epidemiological features of the data
deemed key in \cite{xia2004measles}.

In order to address the above issues, we propose a new approach
that directly focuses on the aspects of the underlying process
that are of scientific interest. We develop a Gaussian process
approximation to the gravity model based on key summary statistics
obtained from simulations of the model at different parameter
values. These statistics are chosen by domain experts to capture
the biologically
important characteristics of the dynamics of the disease. 
The Gaussian process model `emulator' is then used to develop a
probability model for the observations, thereby permitting an
efficient MCMC approach to Bayesian inference for the parameters.
We demonstrate that the new method recovers the true parameters
and the resultant fitted model captures biologically relevant
features of the data.

When applied to the gravity TSIR model, our approach allows us to
investigate several scientific questions that are of interest to
the dynamics of measles. We study changes in dynamics between
school holiday periods versus non-holidays in the pre-vaccination
era. This is particularly interesting because the local,
age-structured transmission rate of the disease changes from
holidays to non-holidays \citep{dietz1976incidence,
bjornstad2002dynamics}. Since our approach allows us to construct
confidence regions easily, we also infer the amounts of exported
and imported infected individuals for different cities during
different time periods and reveal that movement patterns of the
infection do not seem to change significantly between the
pre-vaccination and vaccination eras. Based on the parameter
estimates obtained using our method, we are able to display the
inflow and outflow networks of the infection between cities. Along
with histograms of the degree distributions of the networks, these
graphs help to identify the cities that are important hubs in
measles transmission. More generally, the methodology we develop
here may be useful for models where the likelihood is expensive to
evaluate or in situations where the likelihood is unable to
capture characteristics of the model that are of scientific
interest. We note that the computational cost of forward
simulations for our model makes approaches based on approximate
Bayesian computation (ABC) \citep[cf.][]{pritchard1999population,
marjoram2003markov, beaumont2002approximate} infeasible. Hence our
approach is computationally efficient, while ABC is not a viable
option here.

The rest of the paper is organized as follows. Section
\ref{sec:gravitymodel} describes in detail the gravity TSIR model,
which acts as our motivating example.
 Section \ref{sec:problemandinference} describes the inferential
and computational challenges posed by the model and the large
space-time data set. Section \ref{sec:emulator} describes our new
 emulation-based approach that is an alternative to traditional likelihood-based inference. Section \ref{sec:EmulGravity} describes
computational details and the application of our method to the
gravity TSIR model in simulated data examples. Section
\ref{sec:application} describes the application of our method to
the England-Wales measles data sets. Finally, in Section
\ref{sec:gravdiss}, we summarize our results and discuss our
statistical approach and scientific conclusions.

\section{A gravity model for disease dynamics}\label{sec:gravitymodel}
A general goal of fitting metapopulation disease dynamics models
is to describe spatiotemporal patterns of epidemics at the local
scale and understand how these patterns are affected by the
network of spatial spread of the disease \citep{keeling2004new,
cliff1993measles}. The gravity model we study is an extension of a
discrete time-series susceptible-infected-recovered model
\citep{bjornstad2002dynamics, grenfell2002dynamics} for local
disease dynamics which includes an explicit formulation for the
spatial transmission between different host cities
\citep{xia2004measles}.

The common theoretical framework used to describe the dynamics of
infectious diseases is based on the division of the human host
population into groups containing susceptible, infected
(infectious) and recovered individuals. Let $I_{kt}$ and $S_{kt}$
denote the number of infected and susceptible individuals
respectively in disease generation $t$ in city $k$ and variable
$L_{kt}$ be the number of infected people commuting to city $k$ at
time $t$. The `commuting' assumption reflects that movement of
infection is mostly through transient movement of individuals.
Denote the size and birth rate of city $k$ at time $t$ by $N_{kt}$
and $B_{kt}$, and let $d_{kj}$ represent the distance between
cities $k$ and $j$. The model can then be described as follows.
First, the model for the number of incidences of measles is
\begin{equation}
    \label{eq:eq1}
    I_{k (t+1)}\sim \mbox{Poisson}(\lambda_{k,t+1}) \mbox{, where }
    \lambda_{k,t+1}=\beta_{t}S_{kt}(I_{kt}+L_{kt})^\alpha,
\end{equation} with $t = 1,...,T, k = 1,...,K$, where $K$ is the number of cities in our data and $T$ is
the total number of time steps. The time-step is taken to be 2
weeks, roughly corresponding to the generation length (serial
interval) of measles. The so-called transmission coefficient,
${\bf \beta}:=\{\beta_t\}$, is a parameter that represents the
attack rate of measles at time $t$ and $\alpha$ is a positive real
number correcting for the discrete-time approximation to the
underlying continuous-time epidemic process
\citep{glass2003interpreting}. Since these parameters only affect
the local dynamics of measles, henceforth we refer to these
parameters as "local dynamics parameters." The indexing by $t$ for
$\beta_t$ reflects how this parameter is taken to be a piece-wise
constant taking 26 different values to accommodate seasonal
variability of the transmission rate that is repeated every year
\citep{fine1982measles, finkenstadt2000time,
bjornstad2002dynamics, grenfell2002dynamics}. From this, it can be
seen that $I_{k (t+1)}$ increases depending on the number of
susceptibles and the number of moving infections coming to city
$k$ at the previous time step. Note that we use the Poisson
distribution whereas \cite{xia2004measles} use the Negative
Binomial distribution; this is due to the greater computational
stability of the Poisson distribution for small values of
$\lambda$. Our approach would proceed in the same way for the
Negative Binomial and Poisson assumption. In addition, our
exploratory analysis show that a model fit from using the Poisson
distribution is similar to a model fit obtained with the Negative
Binomial distribution and the final inference about the parameters
of interest is not affected by changing the distributional
assumption.

The susceptibles are modeled as follows
\begin{equation}
    \label{eq:eq2}
    S_{k (t+1)}=S_{k t}+B_{kt}-I_{k (t+1)},
\end{equation} reflecting how susceptibles are replenished by births and
depleted by infection. Since case fatality from measles was very
low for the period of time in this study and mean age of infection
was small, mortalities are not included in this balance equation.
We note that here and in the following, after vaccinations are
available, the birth rates ($B_{kt}$) are deflated by the
corresponding percentage of vaccinated newborns ($V_{kt}$), since
those cannot be infected.

Finally, the gravity model describes the number of moving infected
individuals by
\begin{equation}
    \label{eq:eq3}
    L_{kt}\sim\mbox{Gamma}(m_{kt},1) \mbox{, where }  m_{kt}=\theta
N_{kt}^{\tau_1}\sum\limits_{j=1, j\neq k}^{K}
         \frac{I_{jt}^{\tau_2}}{d_{kj}^{\rho}},
\end{equation} where Gamma(a,b) represents the Gamma distribution with
shape and scale parameters $a$ and $b$ respectively. Here, $b$ is
chosen to be equal to unity based on exploratory analysis of the
fitted model \citep{xia2004measles}. The reason to model immigrant
infection as a continuous random variable lies in the assumption
that the transient infectives do not remain for a full epidemic
generation.

The local dynamics parameters in Equation (\ref{eq:eq1}) have been
estimated previously \citep{bjornstad2002dynamics,
grenfell2002dynamics, finkenstadt2002stochastic}. In this study,
we are interested in learning about the parameters $\theta$,
$\tau_1$, $\tau_2$ and $\rho$ in Equation (\ref{eq:eq3}) as these
parameters control the spatial spread and regional behavior of the
disease. Note, however, that for convenience and numerical
stability, we use a reparametrization of $\theta$, $\theta' =
-\log_{10}(\theta)/5$ throughout the paper.


\section{Parameter inference for the gravity model}\label{sec:problemandinference}

Reliable estimates of the local dynamics parameters $\alpha$ and
${\bf \beta}$ are available for measles dynamics
\citep{bjornstad2002dynamics, grenfell2002dynamics,
finkenstadt2002stochastic, xia2004measles}. Therefore, since we
are only interested in spatial dynamics of the disease, we assume
that these parameters are known and use the estimates obtained
from previous work \citep[cf.][]{xia2004measles} as the true
values. In particular, the local seasonal transmission parameters
for biweeks 1 through 26, $\beta_t$, are taken to be equal to
$\beta_t$ = (1.24, 1.14, 1.16, 1.31, 1.24, 1.12, 1.06, 1.02, 0.94,
0.98, 1.06, 1.08, 0.96, 0.92, 0.92, 0.86, 0.76, 0.63, 0.62, 0.83,
1.13, 1.20, 1.11, 1.02, 1.04, 1.08), and $\alpha$ is assumed to be
0.97. Here, the difference in the values of $\beta_t$ is primarily
related to the fact that attack rates of measles differ depending
on the season of the year since it is known that schools are major
hubs of transmission of the disease. It also known that the true
transmission process is continuous. Since we are considering a
discretized model with a step equal to two week, it is therefore
expected that the true attack rates of the disease could be
higher. This explains the value of $\alpha$ which is slightly less
than unity. In principle, it may be possible to reduce the
dimensionality of $\beta_t$ while still preserving the seasonality
of attack rates of the infection. With lower dimensional
$\beta_t$, one could assume strong priors for the local dynamics
parameters and try to infer these parameters with the remaining
unknown  parameters jointly. However, trying to simultaneously
infer these parameter values still significantly increases the
identifiability issues and further complicates computation.
Crucially, we note that assuming the local dynamics parameters are
known does not have an undue effect on the model fit as has
already been shown in the literature
\citep[cf.][]{xia2004measles}. Assuming the local dynamics
parameters are known leaves us with four unknown parameters,
$\theta'$, $\tau_1$, $\tau_2$ and $\rho$, that we call the gravity
model parameters (in our Gaussian process based approach in
Section \ref{sec:emulator} we will also introduce several other
parameters). In this paper our focus is on investigating the
gravity model parameters and, when possible, obtaining the best
estimates of them with relevant descriptions of their variability.

As suggested by our domain experts, feasible values for the
gravity parameters lie in the interval $[0,2]$ \citep[see
also][]{xia2004measles}. Therefore, we use uniform priors for
$(\theta', \tau_1, \tau_2, \rho)$ in all the inferential
approaches that follow.

The data are spatiotemporal and tend to be high-dimensional,
$546\times952$ in the case of the England-Wales measles data for
the pre-vaccination era and $1326\times354$ for the later time
period (1966-1994). To study whether our fitted model captures
epidemiologically relevant features of the data, we focus on two
important biological characteristics of the process as suggested
by domain experts. These are:

\begin{enumerate}

\item Maximum number of incidences which we will denote by ${\bf M} = (M_1,
\cdots, M_K)$, where $M_i$ is the maximum number of incidences for
the $i$-th city.

\item Proportions of bi-weeks without any cases of infection denoted by ${\bf
P} = (P_1, \cdots, P_K)$, where $P_i$ is the proportion of
incidence free biweeks for the $i$-th city.
\end{enumerate}

An important goal of our work is to find parameter settings (along
with associated uncertainties and dependencies among them) that
yield a model that produces disease dynamics that are as close as
possible to the data in terms of capturing these key properties.

\subsection{A gridded MCMC approach and simulated examples}\label{subsec:likeapproach}

It is easy to see why each evaluation of the likelihood for the
gravity model is expensive. As in many population dynamic models,
the major difficulty is in integrating over high-dimensional
unobserved variables. For our model, $\{L_{kt}\}$ and $\{S_{kt}\}$
are of $K\times T$ dimensions each, which translates to $2 \times
519,792$ in the case of measles data set for the pre-vaccination
era considered in Section \ref{sec:application}. Details of the
likelihood function are given in Web Appendix A.

In this section, using an MCMC algorithm based on the
discretization of a subspace of the parameter space, we describe
 some issues that arise from a traditional likelihood-based
or Bayes approach for inference for the gravity model. Because
likelihood-based inference for the gravity model is
computationally intractable, our gridded MCMC algorithm requires
certain simplifying assumptions and data imputation for
unobservable susceptibles $\{S_{kt}\}$. These assumptions and
details of constructing our gridded MCMC algorithm for parameter
inference are explained in Web Appendix B. We note, however, that
our inferential approach based on a Gaussian process described in
Section \ref{sec:emulator} does not require the simplifying
assumptions, nor does it require data imputation.

We note that all simulated data sets we consider in this work are
generated from the full gravity model described in Section
\ref{sec:gravitymodel} with initial points equal to the actual
observations at $t = 1$. In these examples, the number of
locations, their coordinates, demographic variables, and the
number of time steps are the same as those in the measles data
described in Section \ref{subsec:data}.

\begin{figure}[htbp]
 \begin{center}
  \includegraphics[scale=0.4]{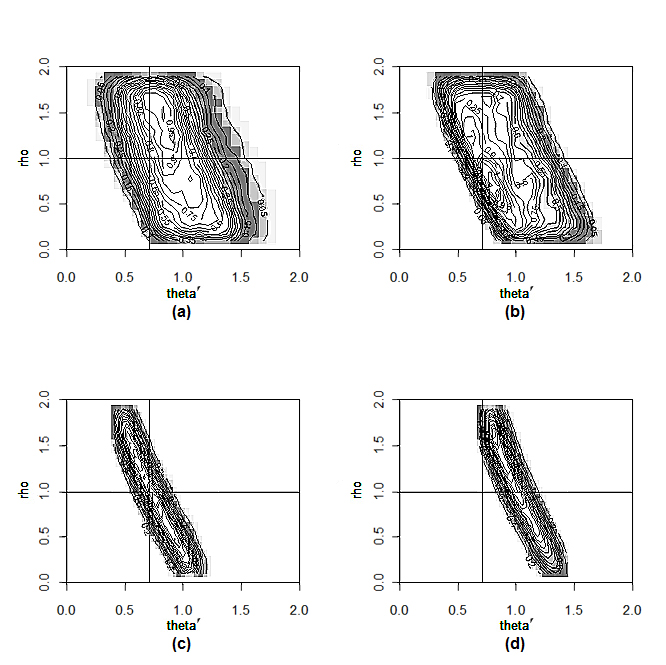}
  \caption{Inferred posterior 2D likelihood surface obtained for data with
known parameters ($\theta' = 0.71$, $\tau_1 = 0.3$, $\tau_2 = 0.7$
and $\rho = 1$):
  (a) Marginal 2D likelihood surface for ($\theta'$,
  $\rho$);
  (b) Marginal 2D likelihood surface for ($\theta'$, $\rho$) assuming $\tau_1 =
  0.3$ (true);
  (c) 2D likelihood surface for ($\theta'$, $\rho$) assuming $\tau_1 =
0.3$ (true) and $\tau_2 =
  0.7$ (true);
  (d) 2D likelihood surface for ($\theta'$, $\rho$) assuming $\tau_1 = 0.5$ (any
value) and $\tau_2 = 1$ (any value).}
  \label{fig:thetavsrho_1}
 \end{center}
\end{figure}

In our first example, we simulate a data set using values for the
gravity parameters $\theta' = 0.71$, $\tau_1 = 0.3$, $\tau_2 =
0.7$ and $\rho = 1$. This parameter setting results in realistic
data that resembles the observations. Figure
\ref{fig:thetavsrho_1} shows conditional and unconditional
posterior likelihood surface plots for $\theta'$ and $\rho$
obtained by using the above gridded MCMC approach. From these
plots, we can easily see that inference for $\theta'$ and $\rho$
is not possible because of the apparent issue with identifiability
(Figure \ref{fig:thetavsrho_1} (a)). In Figure
\ref{fig:thetavsrho_1} (b) we see that identifiability is reduced,
but still exists when we fix one of the parameters, say $\tau_1$,
at its known true value. In Figure \ref{fig:thetavsrho_1} (c), we
fix both of $\tau_1$ and $\tau_2$ at their true values and see
that the obtained ridge contains the true values for $\theta'$ and
$\rho$. Figure \ref{fig:thetavsrho_1} (d) demonstrates that the
ridge moves by changing the values of $\tau_1$ and $\tau_2$ away
from their true values.

\begin{figure}[htbp]
 \begin{center}
  \includegraphics[scale=0.33]{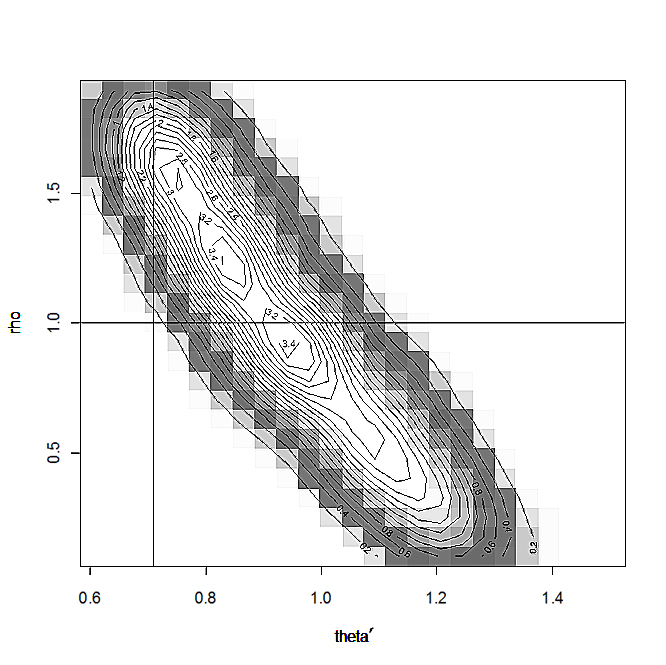}
  \caption{Inferred posterior 2D likelihood surface obtained for data with
known parameters ($\theta' = 0.71$, $\tau_1 = 0.5$, $\tau_2 = 1$
and $\rho = 1$): Posterior 2D likelihood surface for ($\theta'$,
$\rho$) assuming $\tau_1 = 0.5$ (true) and $\tau_2 = 1$ (true) has
a shift and does not contain the true ($\theta'$, $\rho$) at its
highest probability area.}
  \label{fig:thetavsrho}
 \end{center}
\end{figure}

In our second example, we simulate a data set using values for the
gravity parameters $\theta' = 0.71$, $\tau_1 = 0.5$, $\tau_2 = 1$
and $\rho = 1$. Figure \ref{fig:thetavsrho} is a plot of the
two-dimensional likelihood in $\theta'$ and $\rho$ space obtained
by fixing $\tau_1$ and $\tau_2$ at their true values 0.5 and 1
respectively. We can see here that the true values of the
parameters of interest are not in the region where the likelihood
is maximized. This, unfortunately, means that repeating the above
with other simulated data with different true values for the
gravity parameters reveals that the ridge analogous to the ridge
in Figure \ref{fig:thetavsrho_1} (c) does not always have to
contain the true values for $\theta'$ and $\rho$. From our study
of multiple simulated data, we also find that the likelihood ridge
can have an intercept that is different from the ridge that we
would intuitively think as the true ridge while having the same
slope. This difference in intercepts creates a shift thereby
resulting in poor parameter inference. Unfortunately the magnitude
and direction of the shift depends on the true parameter values,
so no simple bias correction is available. At first, one may think
that the discretization of the parameters $\tau_2$ and $\rho$ may
be causing some of these issues. We verify that this is not the
case by simply computing the values of the true likelihood
function at the top of the ridges obtained with the discretized
likelihood. We are able to see that the likelihood surface using
the discretization is similar to the true likelihood surface. The
poor inference from our traditional Bayes approach is therefore
clearly not a result of the discretization.

By generating additional simulations using a simpler model where
we fix all the latent variables at their means we also find the
full gravity model does not substantially differ from the simpler
one in terms of capturing interesting biological characteristics
of the underlying dynamics of the disease. In order to study the
effect of this fixing on the likelihood surface, we save the true
latent variables while simulating data and use them in our gridded
MCMC in place of the expectations used in our gridded MCMC
algorithm. The results show that using the true values of the
latent variables does not change the traditional Bayes inference.
This also confirms that the shifts that we observe in the
traditional Bayes approach are not due to simplifying the model in
gridded MCMC algorithm (see Web Appendix B for details about these
assumptions), but rather due to inherent problems with the
likelihood function.

\begin{figure}
 \begin{center}
  \includegraphics[scale=0.15]{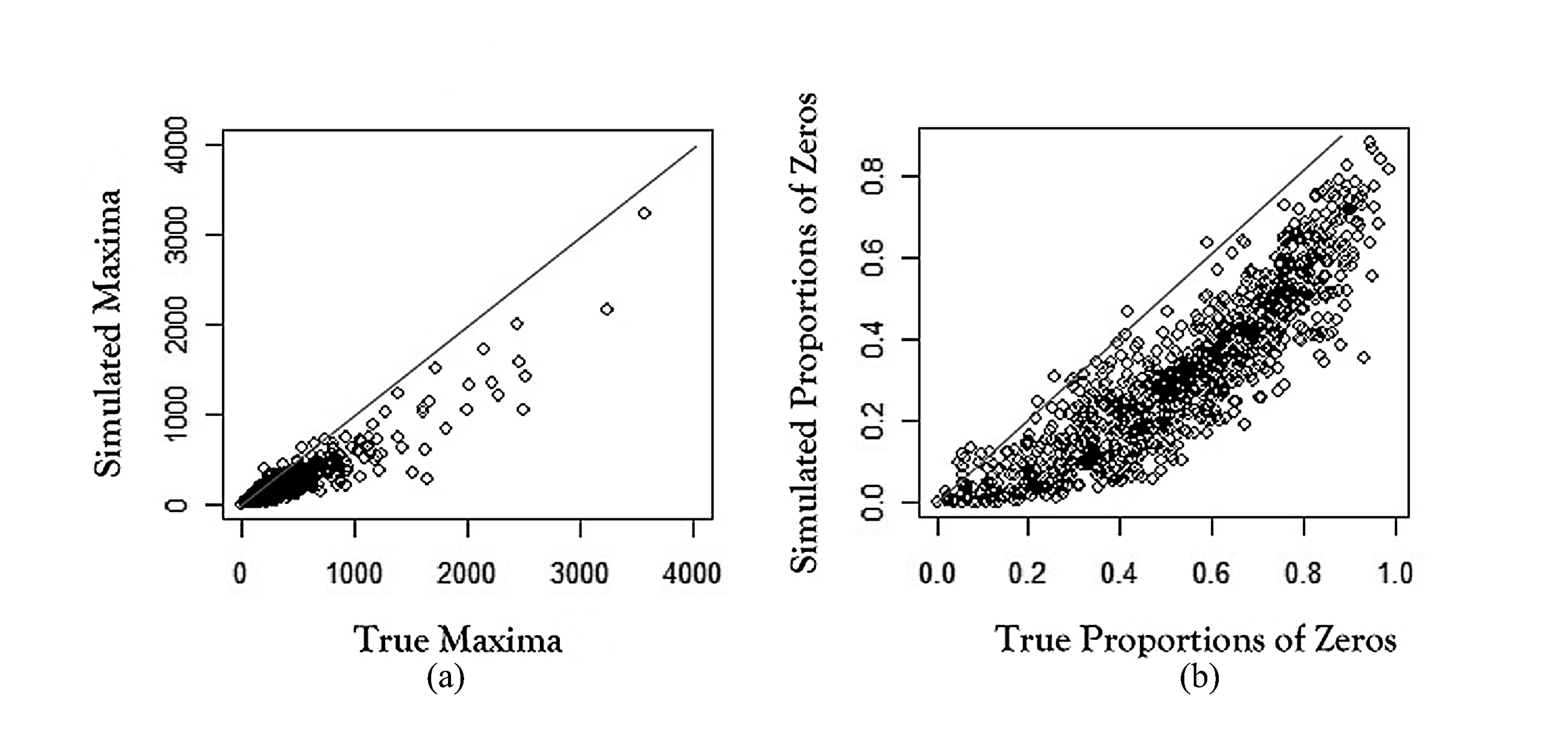}
  \caption{Characteristics of simulated data at the parameters obtained via the
traditional Bayes approach:
   (a) Simulated {\bf M} vs {\bf M} from the data;
   (b) Simulated {\bf P} vs {\bf P} from the data.}
  \label{fig:MLEandBiology}
 \end{center}
\end{figure}

We note that our main interest is to examine whether the parameter
estimates result in a model fit that is capable of reproducing
important characteristics of the observations. In order to study
the model fit from the gridded MCMC, we simulate a data set using
the full gravity model with estimated values of the parameters,
where here and throughout the paper, we use modes of the
corresponding posterior density functions as estimates of the
parameters. These estimates for the measles data described in
Section \ref{subsec:data} are $(\theta', \tau_1, \tau_2, \rho) =
(0.71, 0.5, 1, 1.48)$. For the simulated data set, we calculate
the two 952 dimensional vectors (number of cities in the data) of
summary characteristics and plot them against the summary vectors
for the observed measles data (Figure \ref{fig:MLEandBiology}). We
can see that the simulated data do not seem to match the actual
data in terms of the maximums {\bf{M}} and the proportions of
zeros {\bf P} (Figure
\ref{fig:MLEandBiology} (a)-(b)). 
In Section \ref{subsec:simstudyII}, we compare the model fit
obtained via the gridded MCMC to the model fit we obtain via our
Gaussian process-based approach described in Section
\ref{sec:emulator}.


We summarize below our conclusions based on the gridded MCMC
approach:

\begin{enumerate}
\item The confidence regions for the parameters are very wide,
suggesting that there may be relatively little information even
with a fairly rich data set. Hence we assume that $\tau_1 = 1$,
$\tau_2 = 1$ as estimated in \cite{xia2004measles} and study the
joint distribution of $\theta'$ and $\rho$, which becomes well
informed by the
data. 

\item The fitted gravity model, using the above inference about its
parameters, does not capture important biological features of the
data.

\item We find that the parameter estimates from the traditional
Bayes approach are shifted and the direction of the shift varies
as shown in Figure \ref{fig:thetavsrho}. For example, for a
simulated data set using the parameters values $(\theta', \tau_1,
\tau_2, \rho) = (0.71, 0.5, 1, 1)$, our attempt to infer $\rho$
assuming other parameters are known results in an estimate
$\hat{\rho} = 1.5$ with a confidence region that does not contain
the truth .

\end{enumerate}

\section{Gaussian processes for emulation-based inference}\label{sec:emulator}

Since a traditional Bayes approach suffers from the above
shortcomings, we develop an alternative method that is directly
linked to the characteristics of the infectious disease dynamics
that are of most interest to biologists. This method is based on
using a Gaussian process to emulate the gravity model. A short
review of Gaussian process basics is provided in Web Appendix C.

We describe a new two-stage approach for inferring the gravity
parameters. In the first stage, we simulate the gravity model at
several parameter settings. For each forward simulation of the
model we can calculate the vector of summary statistics based on
the simulated data set. This vector is high-dimensional, 952 (354)
dimensions in the case of measles data for 1944-1965 (1966-1994).
Since Gaussian process-based emulation for high dimensions poses
serious computational challenges, we emulate the model by fitting
a Gaussian process to the Euclidean distances between the summary
statistics of the simulated data at the chosen parameter settings
and the summary statistics for the real data. 
In the second stage, we perform Bayesian inference for the
observations using the GP emulator from the first stage. We also
allow for additional sources of uncertainty such as observational
error and model-data discrepancy as described below. We note that
such two-stage approaches to parameter inference in complex models
has been used to reduce computational challenges and alleviate
identifiability issues \citep[cf.][]{liu2009modularization,
sham2012inferring}.

We begin with some notation. Let $Z$ denote the vector of summary
statistics of interest (e.g. proportions of zeros) calculated
using the observed space-time data set. Let $\Theta$ be the
gravity parameters and $Y(\Theta)$ denote the vector of summary
statistics obtained using a simulation from the gravity model with
the parameter setting $\Theta$. Let $\Omega = (\Theta_1, \cdots,
\Theta_p)$ be a grid on the parameter space. Our first goal is
then to model $\textbf{D} = (D_1, \cdots , D_p)$, where $D_i$ is
the Euclidean distance between $Y(\Theta_i)$ and $Z$ for $i = 1,
\cdots, p$. This is done in the first stage of our approach where
we assume,
\begin{equation}\label{eq:gp}
\textbf{D}|\Omega,{\bf \beta_G},{\bf \xi_G} \sim N(X{\bf
\beta_G},\Sigma({\bf \xi_G}))
\end{equation} Here, $\xi_G =
(\sigma^2_G,\tau^2_G,\phi_G)$ is a vector of parameters that
specify the covariance matrix, and ${\bf \beta_G}$ is a vector of
regression coefficients. The matrix $X$ is a design matrix of
dimension $p \times 5$ with $i$-th row equal to $(1,
\Theta_i^{T})$. In other words, columns of $X$ are the values the
gravity parameters,
 $(\theta,\tau_1, \tau_2, \rho)$, on the selected grid and an intercept. We
use Gaussian covariance matrix, $\Sigma(\xi_G)$, elements of which
are given by,
\begin{align*} (\Sigma(\xi_G))_{ij} = &\mbox{cov}(D_i,
D_j)=\\ & = \begin{cases}
\sigma_G^2\exp(-\phi_G^2||\Theta_i-\Theta_j||^2), &\mbox{if } i\neq j\\
\sigma_G^2+\tau_G^2, &\mbox{otherwise.}
\end{cases}
\end{align*} Here, $||a-b||:=d(a-b,a-b)$, where throughout the paper, the function $d(\cdot,\cdot)$ returns
the Euclidean distance between the argument vectors.  Then, if we
let the maximum likelihood estimate of $({\bf \beta_G},{\bf
\xi_G})$ be $(\hat{\beta}_G,\hat{\xi}_G)$, using standard
multivariate normal theory \citep[cf.][]{Ande:an:1984}, the normal
predictive distribution for the simulated distance $D$ at a new
$\Theta$ can be obtained by substituting
$(\hat{\beta}_G,\hat{\xi}_G)$ in place of $(\beta_G,\xi_G)$ and
conditioning on $\textbf{D}$. We denote this predictive
distribution by $\eta(D;\Theta)$. Detailed version of constructing
this predictive distribution (emulator) is given in Web Appendix
D.

Consider a new space-time data set, and let the vector of summary
statistics for these data be $Y^\ast$. Let the distance between
$Y^\ast$ and $Z$ be $D^\ast$. The predictive distribution from the
first stage provides a model for $D^\ast$,
$\eta(D^\ast;\Theta^\ast)$, connecting it to some unknown
parameter vector $\Theta^\ast$.

Following \cite{bayarri2007framework}, we model the discrepancy
between the gravity model and the real data. Failing to account
for data-model discrepancy can lead to poor inference as pointed
out in \cite{bayarri2007framework} and
\cite{Bhat:Hara:Goes:comp:2010}. We account for this by setting
$D^\ast=D^\ast_\delta:=\delta$, where $\delta>0$ is the
discrepancy term. It is positive since it represents an Euclidean
distance that is non-negative (in the unrealistic case that there
is an exact match between the model for the data and the model
used to fit the data, $\delta$ would be identically equal to 0).
We then infer the gravity parameters using
$\eta(D^\ast_\delta;\Theta^\ast)$ considering $\delta$ to be
another unknown parameter in the MCMC algorithm. In other words,
the likelihood function we use for our MCMC algorithm is a
function $f(\delta,\Theta^\ast):=\eta(D^\ast_\delta;\Theta^\ast)$.
We note that including a model discrepancy term results in more
reliable parameter inference with narrower confidence regions
since it adjusts for the fact that even the best model fit is not
going to reduce the distance between the simulated and observed
summary statistics to zero. In our simulated examples, where data
are generated from the gravity model, the discrepancy term can be
thought of as an adjustment parameter for the fact that two data
sets simulated at the same parameter settings will always have
small differences due to stochasticity. In these examples, as it
is expected, estimate of the discrepancy is very small compared to
the discrepancy term inferred from the original data. We also note
that using negative values for $\delta$ would mean an
extrapolation in our emulator beyond the grid of the parameter
space that may lead to unreliable inference. In many situations,
having a well-defined discrepancy term with an informative prior
helps to reduce problems with identifiability of the parameters as
well \citep[cf.][]{craig2001bayesian}.

We can now summarize our inferential approach as follows:
\begin{enumerate}
\item[1.] Emulating the gravity model:
\begin{enumerate}
\item Select a grid $(\Theta_1, \cdots,
\Theta_p)$ on the range of possible values for $\Theta$.
\item Calculate $Y(\Theta_i)$ using a simulation from the gravity
model with $\Theta_i$ for all $i$.
\item Calculate $\textbf{D} = (D_1, \cdots , D_p)$, distances from
$Y_i$ to $Z$ for all $i$.
\item Find the maximum likelihood estimates of $({\bf \beta_G},{\bf \xi_G})$, the
parameters of the Gaussian process in Equation (\ref{eq:gp}).
Obtain the predictive distribution $\eta(D;\Theta)$.
\end{enumerate}
\item[2.] Bayesian inference for $\delta$ and $\Theta^\ast$ given the
observations  $Z$:
\begin{enumerate}
\item Using the predictive distribution with a discrepancy term,
$\eta(D^\ast_\delta; \Theta^\ast)$, perform Bayesian inference for
the parameters $(\Theta^\ast, \delta)$ from the posterior
distribution via MCMC.
\end{enumerate}
\end{enumerate}

\section{Emulation-based inference for the gravity TSIR model}\label{sec:EmulGravity}

In this section we describe details of the application of the
inferential approach described in Section \ref{sec:emulator} to
the gravity TSIR model. By using simulated data examples, we show
that the approach resolves the problems posed by traditional
approaches. In order to contrast our approach to a traditional
likelihood-based approach (carried out by gridded MCMC as
described in Section \ref{subsec:likeapproach}), we also provide
computational details from the application of both methods.

\subsection{Computational details of gridded MCMC and emulation-based approaches} \label{subsec:comp_details}

Inference for both the traditional Bayes and emulator-based
approaches relies on sampling from the corresponding posterior
distributions via MCMC. In both methods, we use univariate
sequential slice sampling updates for the continuous parameters
\citep{neal2003slice, agarwal2005slice}. Parameters that are on
the grid are updated via an analog of a simple random walk for
discrete variables. In all the MCMC algorithms that are used for
the discretized MCMC approach, the chain is run until we obtain
200,000 samples. This takes about 3 days on a Intel Xeon E5472
Quad-Core 3.0 GHz processor. In all the MCMC algorithms for the
Gaussian process-based method, all the updates are carried out
using slice sampling since all the parameters here are continuous.
Chain lengths are 200,000 again and it takes about 10 hours to
generate them. The chain lengths in both methods are adequate for
producing posterior estimates with small Monte Carlo standard
errors \citep{flegal2008markov, jones2006fixed}.

We emulate the gravity model with a Gaussian process using
proportions of zeros as a summary statistic of interest. Our
selection of proportions of zeros as the primary summary statistic
of the analysis is based on suggestions by domain experts and
intuition that these summary statistics are the most informative
regarding the parameters of interest. It could be argued that big
cities do not have bi-weeks without incidences of measles making
the proportions of zeros for these cities equal to 1. However,
during the course of outbreaks in these cities, the epidemic
trajectory of measles is nearly unaffected by infection that may
enter from neighboring locations. This means that big cities may
not contain information about the gravity parameters - parameters
of the movement of the infection between cities from data on
number of cases of measles. In our data, more than $90\%$ of the
cities may be considered as small cities. Spatial transmission is
very important to the dynamics of measles for these smaller cities
where the infection may become locally extinct. For small cities,
infection re-entered from other cities is the only possible way to
start a new outbreak.

Using different summary statistics may, of course, lead to
different inference. Inference based on the maximums, however, was
identical to what is obtained here and therefore we do not include
details of the analysis and the corresponding results. It is also
possible to develop an emulator using these two summary statistics
at the same time; this is computationally more demanding and based
on our exploratory data analysis will not impact our conclusions.

In general the most informative summary statistics are not trivial
to judge, and depend on the disease and available data. The choice
of summary statistics is closely linked to the particular
inference questions addressed and can be limited by the
availability of informative statistics for any particular model
parameters. In cases when there are no well-established summary
statistics and/or scientifically important aspects of the disease
dynamics that need to be captured, our emulation-based approach
can be used with summary statistics constructed/selected via
algorithms borrowed from the approximate Bayesian computation
literature \citep[cf.][]{fearnhead2012constructing, blum2010non,
nunes2010optimal, sisson2010likelihood}. A possible approach to
the lack of informative summary statistics is to increase the
number of summary statistics, thereby hoping to increase the
amount of information regarding the unknown parameters
\citep{sousa2009approximate}. This approach could, however, make
our inferential methods more computationally expensive. Another
method for selecting summary statistics is based on ordering
summary statistics according to whether their inclusion in the
analysis substantially improves the quality of inference defined
by different criteria \citep{nunes2010optimal,
joyce2008approximately}. Finally, one may construct informative
summary statistics using different dimension reduction techniques
\citep{wegmann2009efficient, fearnhead2012constructing,
blum2010non} or by transforming the existing summary statistics
\citep{blum2010approximate}.

We use the priors for the gravity model parameters that are
described in Section \ref{sec:problemandinference}. Since the
discrepancy term, $\delta$, is always positive, we use an
$\text{exponential}(1)$ as its prior distribution.

We use a uniform grid in the four-dimensional cube, each side of
which is equal to the intervals $[0, 2]$. For each parameter, we
use 20 different values on each axis of the cube; this grid size
permits computationally expedient inference. Our analysis of
simulated data sets also shows that 20 is sufficient for accurate
inference. In addition, for each point on the grid, the average
distances from multiple forward simulations can be used instead of
the distances calculated from a single simulation. This may be
important when model realizations are highly variable. For the
parameters of the gravity model, however, our inference was
insensitive to the number of repetitions. This was because
multiple realizations from the probability model varied very
little for a given parameter setting. Therefore, it was much more
important to use our computational resources for emulation across
more parameter settings than it was to obtain repeated
realizations at the same setting. Hence, we used one simulated
time-series at each location for each set of parameters in the
four-dimensional cube.

\subsection{Application to simulated data}\label{subsec:simstudyII}

\begin{figure}
 \begin{center}
  \includegraphics[scale=0.6]{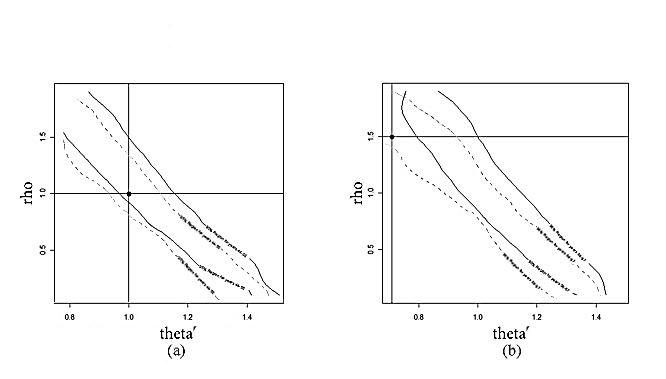}
  \caption{ 95\% C.I.'s for ($\theta'$, $\rho$) obtained via different
  methods (assuming that $\tau_1$ and $\tau_2$ are known):
   Solid line shows the 95\% region obtained using the traditional Bayes method.
Dashed line outlines the 95\%
   region obtained via GP emulator: (a) Both regions contain the true parameter
values; (b) Region obtained by the GP
   emulator contains the true values of the parameters, while the
traditional Bayes region does not.}
  \label{fig:GPonMLE_1}
 \end{center}
\end{figure}

In the simulated examples that follow, our goal is to compare
inference based on the GP-approach to inference from the
traditional Bayes approach. In Figure \ref{fig:GPonMLE_1}, we show
a simulated example where both the GP and traditional Bayes
approaches yield the same inference, and another simulated example
where the two approaches yield different answers. In both cases,
the emulation-based approach provides inference that captures the
true parameter values. In the first simulated data, the true
parameters are $\theta' = 1$, $\tau_1 = 0.6$, $\tau_2 = 1$ and
$\rho = 1$. In Figure \ref{fig:GPonMLE_1} (a), we overlay two
different 95\% confidence regions obtained using the two different
methods. Both of these regions are found by assuming $\tau_1=0.6$
and $\tau_2=1$. We can see that for this example, both solid
(traditional Bayes) and dashed (GP emulator-based) regions contain
the true values of $\theta'$ and $\rho$. This shows that inference
based on the GP emulator is as good as inference based on the
traditional Bayes method. To demonstrate that the new approach is
better than the traditional Bayes approach, we choose a second set
of values for the gravity parameters ($\theta' = 0.71$, $\tau_1 =
0.62$, $\tau_2 = 1$ and $\rho = 1.5$) for which we know inference
based on the traditional Bayes approach to be poor (like in Figure
\ref{fig:thetavsrho}). Figure \ref{fig:GPonMLE_1} (b) shows how
the 95\% confidence region from the traditional Bayes method
(outlined with a solid line) is shifted and does not contain the
truth. The permissible region obtained using the GP emulator
(outlined with a dashed line) has corrected the shift and contains
the true values of the parameters.

\begin{figure}[htbp]
 \begin{center}
  \includegraphics[scale=0.15]{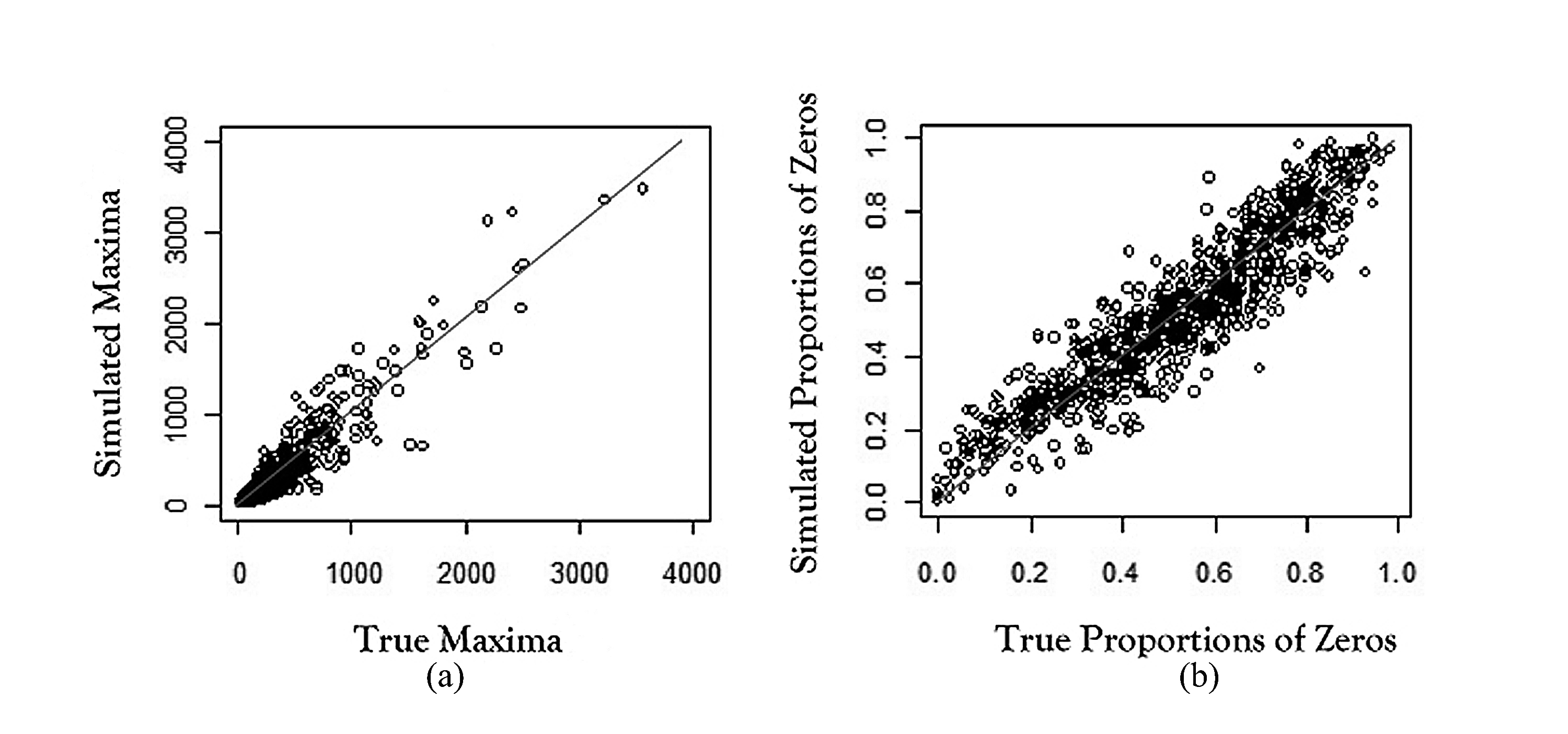}
  \caption{Characteristics of simulated data at the parameters chosen
   to minimize the discrepancy between the data and the simulation:
   (a) Simulated {\bf M} vs {\bf M} from the data;
   (b ) Simulated {\bf P} vs {\bf P} from the data.}
  \label{fig:BESTandBiology}
 \end{center}
\end{figure}

We analyze the ability of the fitted gravity model to reproduce
the key characteristics of the process at these new parameter
estimates. Using estimates obtained via the GP-emulator based
approach, $(\theta', \tau_1, \tau_2, \rho) = (0.71, 0.5, 0.5,
1.48)$, we generate a data set to obtain plots similar to the ones
in Figure \ref{fig:MLEandBiology}. Plots on Figure
\ref{fig:BESTandBiology} (a)-(b) show that the model now can fit
the maximums {\bf M} and the proportions of zeros {\bf P} very
well. Comparing the plots in Figures \ref{fig:MLEandBiology} and
\ref{fig:BESTandBiology}, we can now say that the new
emulation-based approach improves the model fit substantially
while the traditional Bayes parameter estimates from the gridded
MCMC fail to provide a model that captures the key epidemiological
features of the data.

In order to study the effect of a discrepancy term in our
approach, we also tried to infer the gravity parameters using the
emulation-based model with $\delta = 0$ (no discrepancy). The
resultant 95\% confidence regions were much wider for the latter
approach containing incorrect parameter settings, supporting the
points made in \cite{bayarri2007framework} about the importance of
adding a discrepancy term to approximate models. We note, however,
that these new confidence regions still contained the true
parameters values in simulated examples and did not have the kinds
of shifts seen in parameter inference using grid-based MCMC as in
Section 3.1. This means that the problem when the true parameters
of the model are not recovered by a likelihood-based approach is
not related to the issue of accounting for model-data discrepancy.

Continuing to explore the effect of the discrepancy term, we also
tried a few different priors for $\delta$; using the
exponential(1) prior for the discrepancy term worked very well as
was clear from the results. The posterior median for the
discrepancy term was found to be around $2$ which was close to the
minimal distance from the simulated and the true vectors of
summary statistics taken over all the points on the grid.

\section{Results from application to measles data}\label{sec:application}

We apply our emulation-based approach to inference for the gravity
TSIR model to a well known measles data set from the U.K. The
purpose of this is twofold: to demonstrate the applicability of
our approach to a real data set as well as to provide some
insights into measles dynamics in the pre-vaccination era.

\subsection{Description of measles data set}\label{subsec:data}

The following description of the data closely follows
\cite{xia2004measles}. We analyze weekly case reports of measles
for cities in England and Wales. The data is available for K = 952
locations in the pre-vaccination era from 1944 to 1965 and for K =
354 locations from 1966 to 1994 with information on vaccine
coverage. The data represent an interesting case study of
spatiotemporal epidemic dynamics \citep{grenfell2002dynamics} with
well understood underreporting rate of 40$\%$-55$\%$
\citep{bjornstad2002dynamics}. Besides the under-reporting, the
data are complete and reveal inter-annual outbreaks of infection.
A critical feature of this data set is that, except for a few
large cities, infection frequently goes locally extinct, so that
overall persistence hinges on episodic reintroduction and spatial
coupling. Before further analysis, we correct the reported data by
a factor of 1/0.52, with 52\% being the average reporting rate
taken from previous analysis \citep{bjornstad2002dynamics,
clarkson1985efficiency, finkenstadt2000time}. In addition, as in
previous works, we use a timescale that represent the exposed and
infectious period, which is known to be about 2 weeks for measles
\citep{black1989measles}.

In the analysis of the data for pre-vaccination era, following a
standard assumption in the literature \citep[see, for
instance,][and the references
therein]{xia2004measles,bjornstad2002dynamics,
grenfell2002dynamics}, the population sizes and per capita birth
rates for all locations in this work are assumed to be
approximately constant throughout the time period. These variables
are taken as those in 1960 for each of the areas. This is a rough
approximation, since most communities grew during the period we
analyze. The force of infection is, therefore, on average slightly
underestimated (overestimated) during the early (late) part of the
study. In the analysis of the newer data for 1966-1994, the
population sizes and per capita birth rates are allowed to be
variable as specified in the gravity model. We note that these
assumptions are made for the consistency of our work with the
previous analysis and do not have an effect on our inference
and/or conclusions.

\subsection{Some implications for measles dynamics}\label{subsec:measles analysis}

Important biological questions we want to answer based on these
data are: (i) do the gravity model parameters (and hence disease
transmission) change for school holiday periods versus non-holiday
periods? Do they change for different time periods (before and
after vaccines against measles were available)? (ii) do movement
rates of infected people change in different time periods?

\begin{figure}[htbp]
 \begin{center}
  \includegraphics[scale=0.13]{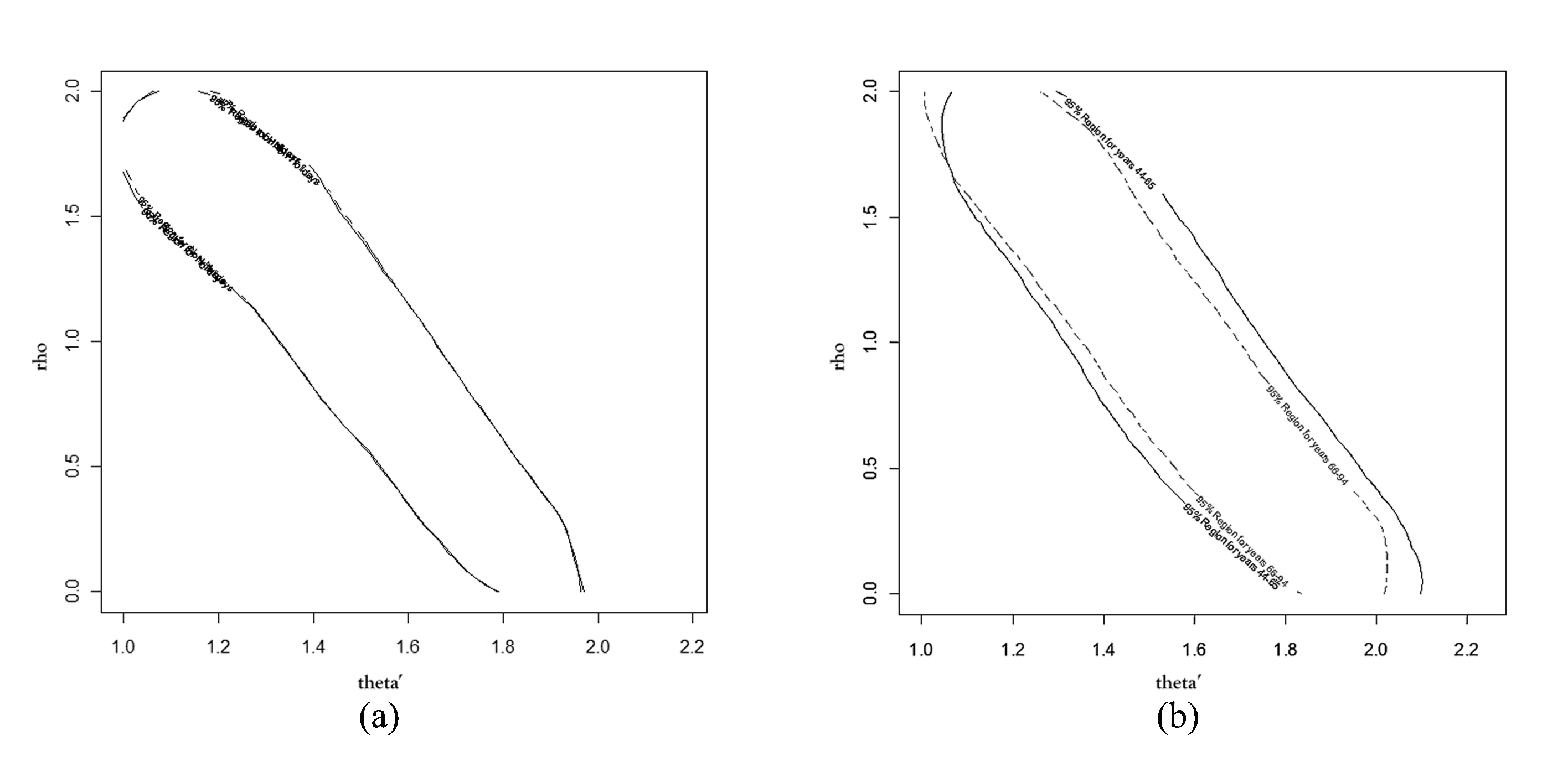}
  \caption{ 95\% C.I. for ($\theta'$, $\rho$) obtained via fitting GP emulator
to a part of the data: (a) Solid line outlines the confidence
region for parameters when data from only holiday periods are
used; Dashed line outlines the confidence region for parameters
when data for only non-holiday periods are used; (b) Solid line
outlines the confidence region for parameters when data for years
from 1944 -1965 are used; Dashed line outlines the confidence
region for parameters when data for years from 1966 - 1994 are
used.}
  \label{fig:GPon4455}
 \end{center}
\end{figure}

In order to answer these questions, using our emulator-based
approach, we first fit the model to the parts of the data
corresponding to periods of holidays and non-holidays. As
demonstrated in our simulated examples in Section
\ref{subsec:likeapproach} and \ref{subsec:simstudyII}, it is not
possible to infer all the gravity model parameters at once. Hence,
we set the parameters $\tau_1$ and $\tau_2$ equal to 1 and study
the remaining key gravity model parameters $\theta$ and $\rho$.
The resulting 95\% confidence regions for $\theta'$ and $\rho$ are
provided in Figure \ref{fig:GPon4455} (a). As can be seen from
this figure, the two regions are almost identical, indicating that
any change in the number of cases of measles for holidays and
non-holidays is not due to the change in the way the infection
spreads between cities of the metapopulation during these periods.

Since the matrix $ M =\{m_{kj}\}$, where $m_{kj}=\theta'
N_{kt}^{\tau_1}\sum\limits_{t=1}^{T}\frac{(I{jt})^{\tau_2}}{d_{kj}^{\rho}}$
is interpreted as a matrix of the amount of movement, sum of
$k$-th row of $M$ represents the amount of infected individuals
leaving city $k$ while sum of $k$-th column is the number of
infected people coming to city $k$. Using samples for $\theta'$
and $\rho$, we easily obtain a sample for the spatial flux of
infection for selected cities. In Table 1, we report our estimates
with corresponding credible regions based on this analysis. We use
the posterior median as point estimates. For example, we estimate
the average number of emigrating infections during the holiday
periods each week to be equal to 31.1 for London. Below the
estimate, we report a 95\% credible interval for it which is (4.4,
479.1). Based on these estimates, the mobility of the infection
appears to be less during the periods of holidays.

\begin{table}
\caption{Estimated amount of average movement in two
weeks}\centering
\begin{tabular*}{0.75\textwidth}{c l l l l}\hline
    \multirow{2}{*}{City}       &\multicolumn{2}{c}{From} &
\multicolumn{2}{c}{To} \\ \cline{2-5}
                                & Holiday          & Non-Holiday       &
Holiday           & Non-Holiday         \\ \hline
    \multirow{2}{*}{London}     & 31.1             & 46.6              & 34.4
           & 49.8                \\
                                & (4.4, 479.1)     & (6.6, 744.7)      & (4.6,
564.9)      & (6.9, 823.9)        \\
    \multirow{2}{*}{Birmingham} & 7.5              & 10.8              & 7.5
           & 11.5                \\
                                & (1.2, 72.9)      & (1.8, 110.6)      & (1.2,
74.7)       & (1.9, 115.8)        \\
    \multirow{2}{*}{Manchester} & 7.8              & 10.3              & 9.1
           & 10.8                \\
                                & (1.0, 151.4)     & (1.4, 180.9)      & (1.2,
162.9)      & (1.5, 189.1)        \\
    \multirow{2}{*}{Blackpool}  & 0.8              & 1.1               & 0.6
           & 0.7                 \\
                                & (0.1, 6.7)       & (0.2, 8.8)        & (0.1,
5.2)        & (0.1, 6.1)          \\ \hline
\end{tabular*}
\label{table:time_holid}
\end{table}

\begin{figure}[htbp]
 \begin{center}
  \includegraphics[scale=0.14]{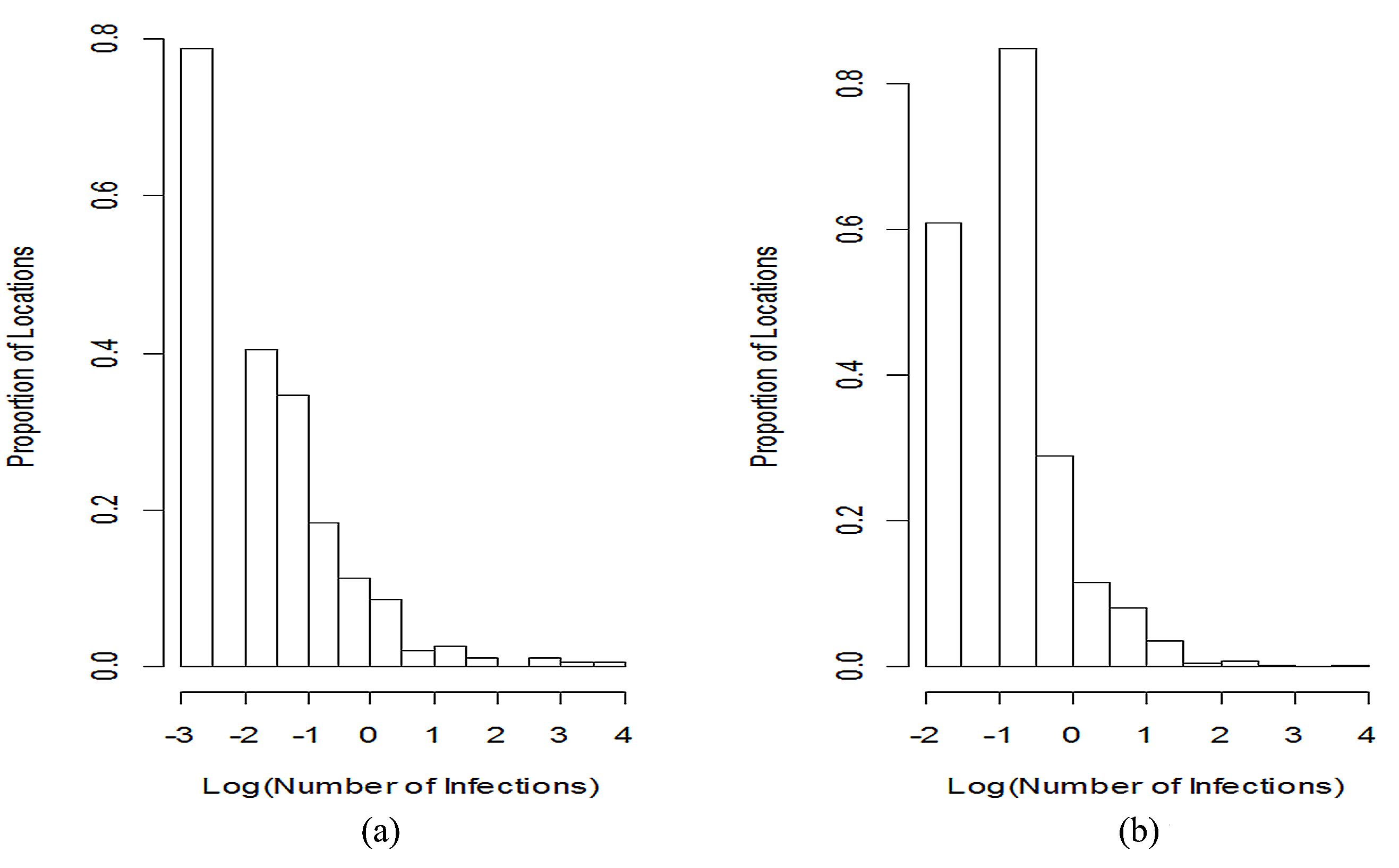}
  \caption{Histogram of estimated amount of average movement in two
weeks for 1966-1994: (a) outgoing infections; (b) incoming
infections.}
  \label{fig:degree}
 \end{center}
\end{figure}

Figure \ref{fig:GPon4455} (b) shows confidence regions obtained by
the GP emulator-based approach by fitting the model to the data
from 1944-1965 and 1966-1994 separately. From this figure, we
conclude that the change in parameter values is statistically
insignificant for these two different time periods. The important
scientific implication of this result is that introduction of
vaccination in England and Wales in 1966 does not change the
movement patterns of the infection between cities. This also means
that any observed change in incidence rates of measles is only due
to the effects of vaccination, not a change in movement patterns
in the vaccination era. Table 2 shows estimates of the average
amount of transit infections each bi-week for years 1944-1965 and
1966-1994. We see here that the infection appears to move less
during the later years. We note that none of the differences are
statistically significant. As a visual summary of this table for
the time period with vaccination, in Figure \ref{fig:degree}, we
plot histograms of log-transformed estimated amount of average
movement in two weeks for 1966-1994. From these plots, we can
conclude that both incoming (Figure \ref{fig:degree} (a)) and
outgoing (Figure \ref{fig:degree} (b)) number of infections for
most of the cities is very small.

\begin{table}
\caption{Estimated amount of average movement in two
weeks}\centering
\begin{tabular*}{0.75\textwidth}{c l l l l}\hline
    \multirow{2}{*}{City}       &\multicolumn{2}{c}{From} &
\multicolumn{2}{c}{To} \\ \cline{2-5}
                                & 1944-65            & 1966-1994           &
1944-65             & 1966-1994          \\ \hline
    \multirow{2}{*}{London}     & 48.1               & 39.2                &
51.6                & 42.4               \\
                                & (7.9, 488.4)       & (4.4, 623.2)        &
(7.0, 591.5)        & (4.9, 739.8)       \\
    \multirow{2}{*}{Birmingham} & 12.9               & 9.3                &
13.5                & 9.1               \\
                                & (1.9, 75.1)        & (1.2, 112.8)        &
(2.8, 93.7)         & (1.2, 121.3)       \\
    \multirow{2}{*}{Manchester} & 12.4               & 10.1                 &
14.0                & 10.1                \\
                                & (2.1, 128.4)       & (0.9, 176.7)        &
(1.9, 163.6)        & (1.4, 193.1)       \\
    \multirow{2}{*}{Blackpool}  & 1.1                & 0.9                 &
0.9                 & 0.8               \\
                                & ( 0.3, 8.1)        & (0.2, 9.7)          &
(0.1, 7.4)          & ( 0.1, 7.1)        \\ \hline
\end{tabular*}
\label{table:time_1944}
\end{table}

\begin{figure}[htbp]
 \begin{center}
  \includegraphics[scale=0.14]{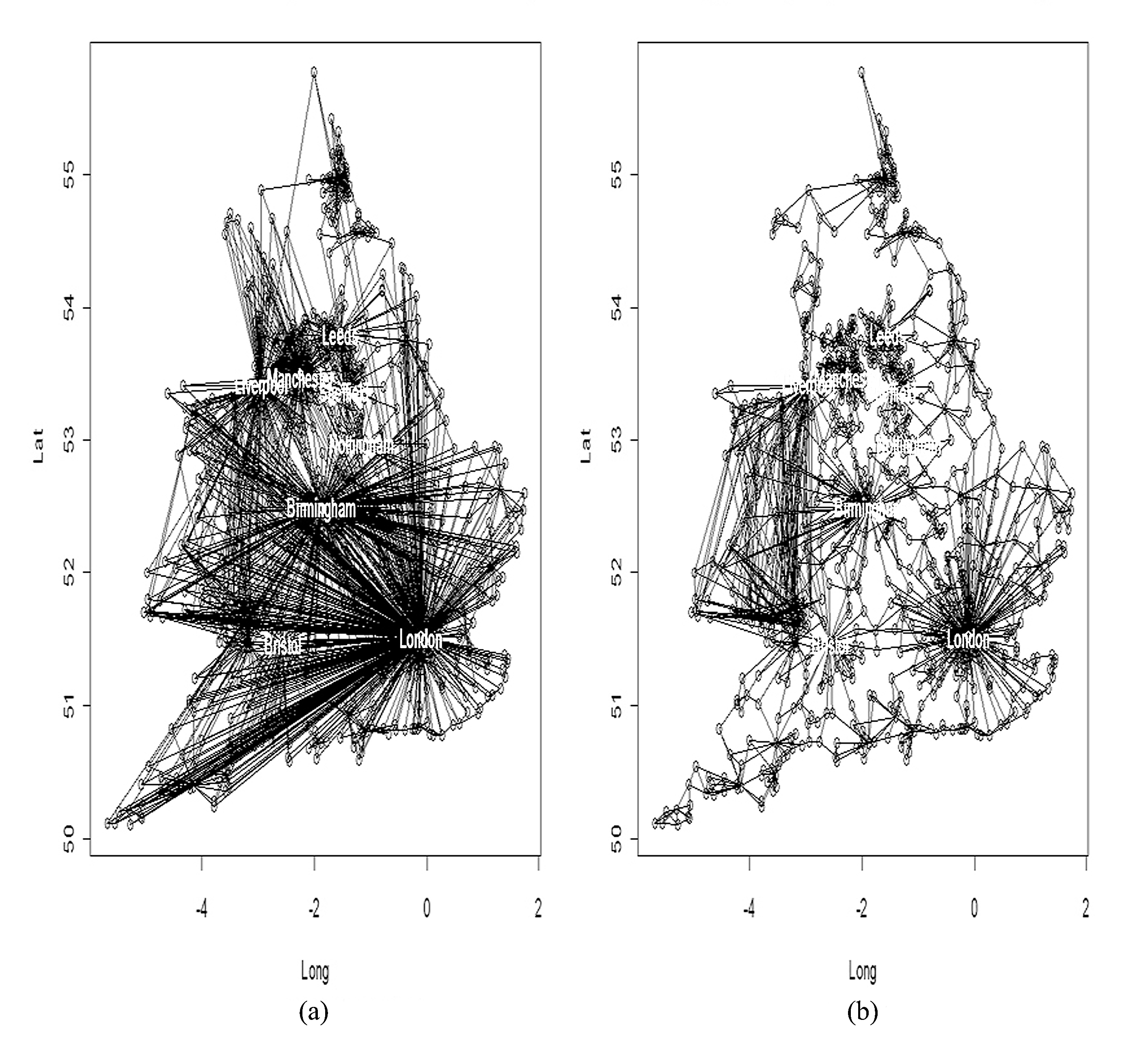}
  \caption{Movement networks of the infection: (a) network of outgoing infections; (b) network of incoming infections.}
  \label{fig:network}
 \end{center}
\end{figure}

Figure \ref{fig:network} displays graphs of networks of the
movement of measles between cities in our data. These graphs are
obtained using the movement matrix $M$ and estimates of the
gravity parameters from data for 1966-1994 via the GP
emulator-based approach. In Figure \ref{fig:network} (a), we plot
the network of outgoing infections. In Figure \ref{fig:network}
(b) we plot the network of incoming infections for cities of the
metapopulation. Figure \ref{fig:network} (a) illustrates the
importance of big cities in the dynamics of measles for smaller
communities where the infection may become locally extinct. From
this figure, we see that the edges radiating from the populated
cities reach the small cities causing a re-introduction of the
infection in these communities. This link between big and small
cities do not seem to depend on distances between the cities. On
the other hand, in Figure \ref{fig:network} (b), we see that the
amount of incoming infections is mostly dependent on distances
between cities since edges connecting different cities in this
graph are shorter relative to the edges of the graph in Figure
\ref{fig:network} (a). This means that big cities are the only
important factors in starting an outbreak in smaller cities,
excluding the possibility of re-introduction of the disease from
neighboring cities with small population sizes.

\section{Discussion}\label{sec:gravdiss}

Complex models are very useful for 
representing physical phenomena, whether the phenomena is the
spread of an infectious disease or the change in sea surface
temperatures in the Atlantic. As is well known, it is not always
possible for every aspect of such complicated phenomena to be
modeled accurately; certain key characteristics of the process
necessarily have to be focal points of the modeling effort.
However, these key characteristics are not typically the focus of
a statistical inferential procedure that uses a traditional
likelihood-based approach. The approach we have developed in this
paper addresses this point by providing a flexible inferential
method that directly takes into account the characteristics of the
process that are most important to scientists. Even though
focusing on different summary statistics can lead to different
estimates, parameter inference based on our approach produces an
improved model fit to the biologically interesting features of the
infectious disease dynamics. In addition to the flexibility this
provides, we find that our approach is also computationally
tractable in situations where traditional likelihood-based
inference is not. In general, when there are no obvious summary
statistics and/or scientifically important aspects of the disease
dynamics that need to be captured and informative for parameter
inference, as was mentioned in Section 5, one can employ our
approach with summary statistics constructed or selected via
existing algorithms in the approximate Bayesian computation
literature \citep[cf.][]{fearnhead2012constructing, blum2010non,
nunes2010optimal, joyce2008approximately, wegmann2009efficient,
blum2010approximate, sisson2010likelihood}.

Computer model emulation and calibration is an active area of
research \cite[cf.][]{kennedy2001bayesian, bayarri2007computer,
sanso2009statistical, rougier2009formal,
Roug:Guil:Maut:Rich:expe:2009, Cont:O'Ha:baye:2010,
Bhat:simu:2007, Higd:Gatt:Will:Righ:comp:2008,
Bhat:Hara:Goes:comp:2010} but most of this work has focused on
deterministic models.

Some authors have also worked on applying the ideas from computer
model emulation and calibration literature in the context
stochastic nonlinear ecological dynamical systems and complex
models is biology \citep[cf.][]{wood2010statistical,
henderson2009bayesian}. The main idea in these papers is to assume
the selected summary statistics are normally distributed random
variables with unknown mean and variance functions that depend on
the parameters of interest of the original model. In the emulation
step, these functions are then estimated from data by utilizing
two different inference algorithms (details are available in
\cite{henderson2009bayesian} and \cite{wood2010statistical}). In
contrast to their approaches, we emulate the scientifically
important summary statistics {\it directly}. That is, we model the
summary statistics, in our case the proportion of zeroes at each
of 952 (or 354) different cities, depending on the data set. This
results in a 952-dimensional summary statistic corresponding to
the observations as well as a 952-dimensional summary statistic
for each simulated data set. While the other approaches may be
easier to implement in cases where the summary statistics may be
assumed to be normally distributed, for the kind of problem we are
considering, our Gaussian process model provides a much more
flexible model for the {\it multivariate} summary statistics. In
the other approaches, a normal model is assumed for the summary
statistics and the Gaussian process model is used to obtain a
flexible model for the means and variances as a function of the
parameters; in fact, this is why their approaches often utilize
transformations of the summary statistics in order to make the
assumption of normality more reasonable. In our approach, a
Gaussian process model (a flexible infinite-dimensional process)
is used to model the summary statistics directly as a function of
the parameters. Hence, our emulation approach captures well
potentially complicated relationships between the parameter and
the multivariate summary statistics, while also providing
uncertainties about this relationship; the uncertainties in the
other approaches describe uncertainties about the mean and
variances of an assumed normal distribution model for the summary
statistics.

Additionally, in our approach we account for the fact that there
is a discrepancy between the output of the biological model and
the data. That is, we do not assume that the biological model
captures the true process perfectly even at some ideal parameter
setting. We allow for an additional process that accounts for
data-model discrepancies. This is important for obtaining
reasonable parameter inference as also pointed out in
\cite{bayarri2007framework} and \cite{sham2012inferring}. In our
view, therefore, our paper makes the following main contributions:
(1) a general inferential approach that focuses on characteristics
(summary statistics) of a process; (2) a method for statistical
inference when the likelihood is intractable {\it and} simulation
from the probability model is expensive; (3) a study of a
particular model for measles dynamics, the gravity TSIR model,
using the approach we have developed.



In the context of measles in the pre-vaccination and vaccination
eras, our method allows us to study some interesting aspects of
the dynamics of measles based on the gravity TSIR model. For
instance, we find that there does not appear to be a significant
change in the gravity parameters for the school holiday periods
versus non-holidays which means that we do not have enough
evidence of a change in the dynamics of measles between these
different periods. By fitting the model using our approach to the
data from 1944-1965 and 1966-1994 separately, we reveal that that
introduction of vaccination in England and Wales in 1966 does not
change the movement patterns of the infection between cities. This
indicates that any observed change in incidence rates of measles
is only due to the effects of vaccination. Analyzing graphs of the
networks of movement of the infection obtained using the estimates
of the gravity parameters from data for 1966-1994, we identify the
important hubs and their roles in transmission of measles between
the cities of the metapopulation. These hubs, the biggest cities,
seem to spread the infection to smaller cities regardless of the
distances between cities, while movement of the infection between
small cities is dependent on the distances.

More generally, the methodology we have described in this paper is
particularly useful in cases where simulation from a probability
model might be too expensive to allow the use of other popular
inferential approaches like ABC. It is worth noting that our
approach works well when the parameter dimensionality is small,
but is generally infeasible for parameter dimensions greater than
around five to eight depending on the model complexity. Our
approach is widely applicable for inference in computationally
expensive but biologically realistic models. In principle,
whenever a likelihood is expensive to evaluate or when traditional
Bayes approaches does not capture the most scientifically relevant
features of the model, our method provides a way to incorporate
important characteristics in a computationally tractable
inferential approach.

\section*{Acknowledgements}

This work was supported in part by a grant from the Bill and
Melinda Gates Foundation.

\bibliographystyle{chicago}
\bibliography{gravitypapers}

\end{document}